# AstMatch: Adversarial Self-training Consistency Framework for Semi-Supervised Medical Image Segmentation

Guanghao Zhu[a], Jing Zhang[a], Juanxiu Liu[a], Xiaohui Du[a], Ruqian Hao[a],
Yong Liu[b], Lin Liu[a,*]

[a] *MOEMIL Laboratory, School of Optoelectronic Science and Engineering, University of Electronic Science and Technology of China, Chengdu, 611731, China*

[b] *School of Optoelectronic Science and Engineering, University of Electronic Science and Technology of China, Chengdu, 611731, China*

**Abstract**

Semi-supervised learning (SSL) has shown considerable potential in medical image segmentation, primarily leveraging consistency regularization and pseudo-labeling. However, many SSL approaches only pay attention to low-level consistency and overlook the significance of pseudo-label reliability. Therefore, in this work, we propose an adversarial self-training consistency framework (AstMatch). Firstly, we design an adversarial consistency regularization (ACR) approach to enhance knowledge transfer and strengthen prediction consistency under varying perturbation intensities. Second, we apply a feature matching loss for adversarial training to incorporate high-level consistency regularization. Additionally, we present the pyramid channel attention (PCA) and efficient channel and spatial attention (ECSA) modules to improve the discriminator's performance. Finally, we propose an adaptive self-training (AST) approach to ensure the pseudo-labels' quality. The proposed AstMatch has been extensively evaluated with cutting-edge SSL methods on three public-available datasets. The experimental results under different labeled ratios indicate that AstMatch outperforms other existing methods, achieving new state-of-the-art performance. Our code will be available at https://github.com/GuanghaoZhu663/AstMatch.

## 1.Introduction

Medical image segmentation can extract important organs or lesions from medical images, which is crucial for biomedical image analysis and surgical planning. Recently, various supervised learning-based encoder-decoder networks have made progress in medical image segmentation tasks, including U-Net [1] and V-Net [2], which largely depend on a significant quantity of annotated data. However, owing to poor contrast and noise interference, labeling medical images is a laborious and expensive manual contouring process. Therefore, it becomes unfeasible to gather substantial datasets of medical images with pixel-level annotations. In contrast to supervised learning, semi-supervised learning (SSL) depends mainly on large amounts of unannotated data with limited amounts of annotated data for effective model training. Consequently, many studies have concentrated on semi-supervised medical image segmentation (SSMIS) [3], [4], [5], [6], [7].

---

[*] *Corresponding Author:* Lin Liu (liulin1979@uestc.edu.cn)

Consistency regularization [8], [9], [3] and pseudo-labeling [10], [11], [12] are the two main methods for image segmentation with SSL. Consistency regularization-based approaches, including Π model [8] and Mean Teacher [9], encourage prediction consistency under different perturbations to mitigate over-fitting in SSL. For these approaches, it is crucial to generate sufficient prediction divergence on unannotated data [13]. Pseudo-labeling-based methods attempt to produce reliable pseudo-labels for the supervised training process of unlabeled samples [14]. The pseudo-labels' quality has a major impact on the effectiveness of these methods [5]. Currently, state-of-the-art (SOTA) approaches like FixMatch [10] have combined consistency regularization and pseudo-labeling. Despite their promising performance, the key factors of these two methods influence each other. Intensifying the degree of prediction disagreement between the sub-networks through strong data perturbations will inevitably increase the uncertainty of pseudo-labels [13]. Therefore, our main goal is to concurrently improve the prediction disagreement and pseudo-label's quality in a unified network for SSL.

In this study, we propose an adversarial self-training consistency framework (AstMatch) based on consistency learning, adversarial learning, and self-training. The motivation is to effectively choose reliable pseudo-labels for self-training by introducing two discriminator networks in a FixMatch framework [10]. Furthermore, we present an adaptive self-training strategy that can fully leverage pseudo-labels identified as both high-quality and low-quality. Our major contributions are outlined as follows:

(1) An adversarial consistency regularization (ACR) approach employing two discriminators is proposed to strengthen weak-to-strong consistency learning. The first discriminator identifies high-quality pseudo-labels, facilitating the knowledge transfer from annotated to unannotated data. The second one distinguishes the segmentation predictions on the same image under strong and weak data augmentations, enhancing consistency learning under varying degrees of perturbations.

(2) A feature matching loss is incorporated to increase the adversarial training's stability. To enhance the discriminator's capacity to identify high-quality pseudo-labels, the efficient channel and spatial attention (ECSA) and pyramid channel attention (PCA) modules are introduced into the discriminator.

(3) An adaptive self-training (AST) strategy is proposed to separately tackle the pseudo-labels with varying qualities identified by the discriminator. High-quality pseudo-labels are directly utilized for self-training, whereas low-quality ones require bidirectional CutMix with labeled data before being used for self-training. The AST strategy effectively ensures the quality of pseudo-labels involved in self-training, enabling continuous improvement in the segmentation network's performance.

We evaluate the performance of AstMatch in three widely used medical image segmentation datasets: ACDC [15], LA [16], and Pancreas-NIH [17] datasets. Extensive experiments demonstrate that our proposed AstMatch outperforms several SOTA approaches, achieving higher accuracy. It is noteworthy that our AstMatch does not introduce new modules to the segmentation network, rendering it a strong baseline for SSMIS.

## 2. Related Work

### 2.1. Semi-Supervised Medical Image Segmentation

Since acquiring substantial quantities of annotated medical images is challenging, many studies have been conducted in SSMIS. Deep learning-based semi-supervised approaches exhibit powerful feature representation capabilities [18], enabling them to strengthen models by extracting semantic features from unlabeled images. To exploit unlabeled data effectively, recent studies focus on incorporating more complicated techniques, such as additional transformer-based branch [19] and consistency regularization between extra decoders [5]. In this study, we concentrate on utilizing consistency regularization, adversarial learning, and self-training to enhance network performance.

Consistency regularization encourages the network to remain constant predictions under different data augmentations [20]. The Mean Teacher (MT) [9] framework is one of the most typical methods, which involves averaging model weights. In the MT framework, after the student model weights have been updated, an exponential moving average (EMA) of them is utilized to update the teacher model weights. The teacher then improves the student with a consistency cost calculated between their predictions on unlabeled data. The MT framework has become a widely used framework for consistency learning, and many works extended it in different ways. For instance, Yu et al. [3] proposed an uncertainty-aware mean teacher (UA-MT) framework, enabling the student model to learn from high-confidence regions based on uncertainty estimation using the Monte Carlo Dropout [21]. Bai et al. [6] emphasized the benefits of a uniform learning process for both annotated and unannotated data in reducing the empirical distribution gap and developed a bidirectional copy-paste (BCP) between them. FixMatch, proposed by Sohn et al. [10], is another commonly used framework for consistency learning, which incorporates pseudo-labeling. Specifically, FixMatch uses high-confidence predicted results on weakly-augmented unannotated data to produce pseudo-labels, which then supervise the model's predictions on the version with strong augmentation to form consistency regularization.

Adversarial learning [4], [7] is an effective approach for extracting latent information from unlabeled data. For example, to establish a geometric constraint with abundant unlabeled data, Li et al. [4] proposed a shape-aware network called SASSNet. In addition to predicting segmentation masks, this network concurrently predicts signed distance maps (SDM). The SDMs of annotated and unannotated data are utilized to calculate the adversarial loss, enforcing a shape constraint for the output. Lei et al. [7] presented an adversarial self-ensembling network (ASE-Net). The framework utilizes two discriminators, with one differentiating the prediction quality disparity between labeled and unlabeled data. Meanwhile, the other discriminator is employed to distinguish between perturbed and unperturbed unlabeled data.

Self-training, a technique that involves generating artificial labels for unlabeled data [10], has become popular in SSMIS. For example, Shi et al. [5] presented a conservative-radical network (CoraNet) with a novel uncertainty estimation method, which utilized the cross-entropy losses with varying misclassification costs to obtain the regions of certainty and uncertainty. The regions of certainty in the prediction are employed as pseudo-labels. For regions of uncertainty, the mean teacher model and a consistent loss are incorporated to ensure reliable pseudo-label assignment.

*2.2. Generative Adversarial Networks*

Goodfellow et al. [22] presented the generative adversarial network (GAN) initially in 2014, which involves two networks: a generative model G that produces samples by passing random noise and a discriminative model D that evaluates the likelihood that a sample originating from the dataset or from G. D is trained to accurately differentiate the training data from samples generated by G, whereas G aims to increase the likelihood that D would make a mistake. This method can be considered as a minimax two-player game. However, due to issues like unstable training and mode collapse in the original GAN [23], many optimized schemes [24], [25], [26] have been proposed. Mirza et al. [24] introduced a conditional GAN (CGAN), extending the GAN into a conditional model by incorporating auxiliary information into the discriminator and generator, thereby enhancing the data generation process. Alec et al. [25] proposed a deep convolutional GAN (DCGAN) capable of learning a hierarchy of representations of images, positioning it as a strong candidate for unsupervised learning tasks. Arjovsky et al. [26] presented a Wasserstein GAN (WGAN), whose optimization function utilizes the Wasserstein distance to increase the training stability.

In medical image segmentation tasks, GANs [27], [28], [29] have been proven to be significant. The generator is often employed to produce the predicted mask, while the discriminator is trained to distinguish it from ground truth. Training the two networks alternately can improve the segmentation performance. Tan et al. [27] presented a GAN-based lung segmentation framework, named LGAN. The generator predicts lung masks and the discriminator calculates the Earth Mover (EM) distance between the masks and ground truth to improve the segmentation accuracy. Li et al. [28] implemented a generative adversarial semi-supervised network, wherein the generator produces pseudo-labels, and the discriminator evaluates their confidence. The reliable pseudo-labels are then utilized to optimize the generator. Tyagi et al. [29] presented CSE-GAN for the segmentation of lung nodules. This architecture incorporates a generator with concurrent squeeze and excitation (CSE) modules [30] and a discriminator with spatial squeeze and channel excitation (sScE) modules [31], improving segmentation and classification accuracy.

Different from the above methods, we first take advantage of the simplicity of FixMatch and employ it as the fundamental framework for our consistency regularization. Second, to identify high-quality predictions as pseudo-labels and enhance consistency learning, we propose a discriminative model incorporating efficient channel and spatial attention (ECSA) modules and a pyramid channel attention (PCA) module. A feature matching loss is employed for adversarial learning, which is essential for the stable training of GAN in SSL. Moreover, we present an adaptive self-training strategy determined by discriminator output scores to make better use of pseudo-labels with varying quality.

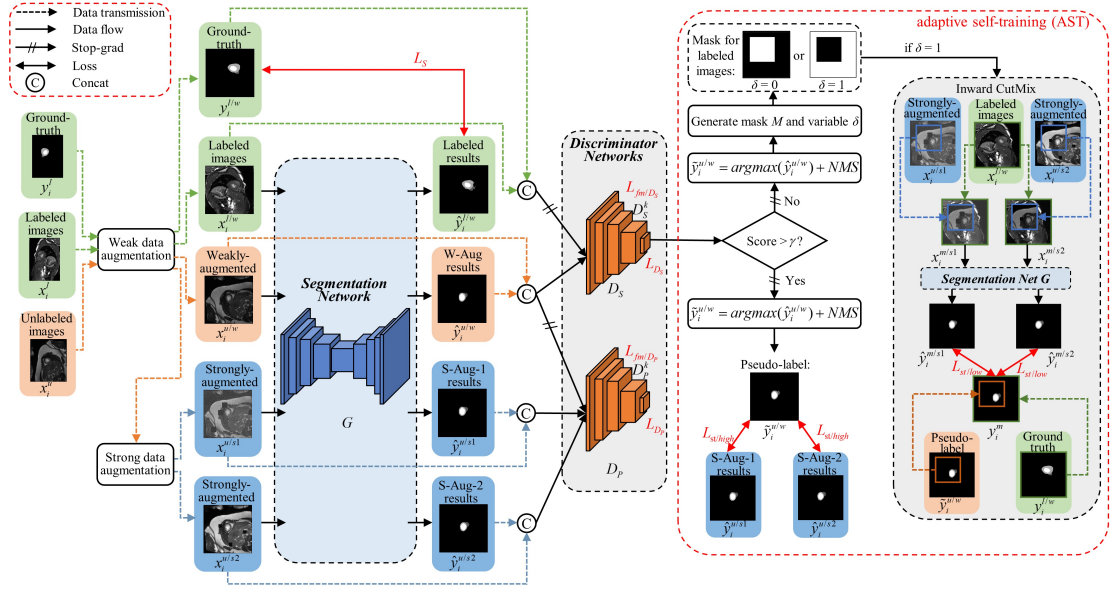

**Fig. 1.** The main framework of our proposed AstMatch. The AstMatch is composed of two main components: the adversarial consistency regularization (left) and the adaptive self-training (right).

## 3. Method

Fig. 1 illustrates our adversarial self-training consistency framework (AstMatch), which is composed of a segmentation network and two discriminator networks. Mathematically, the training set for a semi-supervised segmentation task includes $N$ labeled samples and $M$ unlabeled samples ($N \ll M$). The labeled set is denoted as $D_L = \{(x_i^l, y_i^l)\}_{i=1}^{N}$ and the unlabeled set as $D_U = \{x_i^u\}_{i=N+1}^{N+M}$. $x_i^l \in \mathbb{R}^{H \times W \times D}$ is the labeled input volume with ground-truth annotations $y_i^l \in \{0,1,...,K-1\}^{H \times W \times D}$ and $x_i^u \in \mathbb{R}^{H \times W \times D}$ is the unlabeled input volume. $K$ is the class number.

### 3.1. Adversarial Consistency Regularization

Despite the widespread use of consistency learning in SSL, there are still some issues with it, which are shown as follows:

(1) The Monte Carlo Dropout (MC Dropout) is often employed in SSL to estimate uncertainty [3], [32], [33], [34], [35] and reflect the inaccurately segmented regions in the segmentation results. However, MC Dropout involves multiple forward propagations, leading to a significant increase in training time.

(2) FixMatch adopts a pre-defined constant threshold for the predictions of the weakly-augmented

unannotated data to select pseudo-labels that benefit the training, without considering different learning statuses [11].

(3) Many image segmentation approaches with SSL only focus on low-level consistency and apply consistency regularization at the pixel level, ignoring high-level consistency [36].

To address these problems, we have proposed an adversarial consistency regularization (ACR) approach, extending the FixMatch framework to transfer the latent information from annotated to unannotated data. Firstly, we apply geometrical augmentations (random rotation, flipping, and cropping) to labeled data $(x_i^l, y_i^l)$ and unlabeled image $x_i^u$, generating weakly-augmented $(x_i^{l/w}, y_i^{l/w})$ and $x_i^{u/w}$. Inspired by the dual-stream perturbation framework (DusPerb) in UniMatch [37], we independently generate two strongly-augmented versions $(x_i^{u/s1}, x_i^{u/s2})$ by performing additional RandAugment [38] and Cutout [39] operations on $x_i^{u/w}$. The RandAugment consists of a series of intensity-based augmentations such as contrast enhancement, Gaussian noise, and Gaussian blurring. In the Cutout operation, we randomly select some pixels as center points and place square cutout masks around these locations. Subsequently, we feed the $x_i^{l/w}$, $x_i^{u/w}$, $x_i^{u/s1}$, and $x_i^{u/s2}$ into the segmentation network $G$ to produce corresponding predictions $\hat{y}_i^{l/w}$, $\hat{y}_i^{u/w}$, $\hat{y}_i^{u/s1}$, and $\hat{y}_i^{u/s2}$. The supervision loss $L_S$ is composed of the cross-entropy loss $L_{CE}$ and Dice loss $L_{Dice}$:

$$L_S(\theta_G) = \frac{1}{N}\sum_{i=1}^{N}[L_{CE}(\hat{y}_i^{l/w}, y_i^{l/w}) + L_{Dice}(\hat{y}_i^{l/w}, y_i^{l/w})]/2, \tag{1}$$

where $\theta_G$ represents the segmentation model's parameters; $N$ represents the number of labeled data.

Two discriminators sharing an identical structure are used for distinct purposes. We concatenate the softmax predictions and images into our discriminator networks. Specifically, the segmentation discriminator $D_S$ takes $(y_i^{l/w} © x_i^{l/w})$ and $(\hat{y}_i^{u/w} © x_i^{u/w})$ as inputs, where © represents concatenation along the channel dimension. The $D_S$ learns variations between the segmentation result of unlabeled data and the ground truth of labeled data, which is then utilized for the selection of pseudo-labels with high quality in self-training. In contrast, the perturbation discriminator $D_P$ takes $(\hat{y}_i^{u/w} © x_i^{u/w})$, $(\hat{y}_i^{u/s1} © x_i^{u/s1})$, and $(\hat{y}_i^{u/s2} © x_i^{u/s2})$ as inputs, discerning the differences between the same unlabeled data's predictions under weak and strong data perturbations. Throughout the adversarial training process with the $D_P$, the $G$ learns the prediction consistency of a given unlabeled image under various data augmentation intensities. The loss functions of $D_S$ and $D_P$ are defined as:

$$L_{D_S}(\theta_{D_S}) = \frac{1}{M}\sum_{i=1}^{M}[L_{BCE}(D_S(y_i^{l/w}\copyright x_i^{l/w}),1) + L_{BCE}(D_S(\hat{y}_i^{u/w}\copyright x_i^{u/w}),0)]/2, \quad (2)$$

$$L_{D_P}(\theta_{D_P}) = \frac{1}{M}\sum_{i=1}^{M}[L_{BCE}(D_P(\hat{y}_i^{u/w}\copyright x_i^{u/w}),1) + L_{BCE}(D_P(\hat{y}_i^{u/s1}\copyright x_i^{u/s1}),0) \\ + L_{BCE}(D_P(\hat{y}_i^{u/s2}\copyright x_i^{u/s2}),0)]/3, \quad (3)$$

where $\theta_{D_S}$ and $\theta_{D_P}$ represent the parameters of $D_S$ and $D_P$; and $L_{BCE}(\cdot)$ is binary-class cross-entropy loss.

The segmentation network and discriminators are alternately trained against one another. When the segmentation discriminator $D_S$ fails to differentiate the prediction from ground truth, the segmentation network $G$ exhibits great performance for unlabeled data. When the perturbation discriminator $D_P$ is unable to distinguish between predictions under weak and strong data perturbations, $G$ has predictive consistency under varying perturbation intensities. In summary, this adversarial consistency regularization approach effectively makes use of labeled data to improve the segmentation quality of unlabeled data.

*3.2. Improved Adversarial Loss and Discriminator*

*3.2.1. Improved Adversarial Loss*

Due to the limited availability of labeled data in SSL, ensuring the stability of GAN training is crucial. To enhance this stability and take high-level consistency into consideration, we incorporate two feature matching (FM) losses [23]. The first FM loss $L_{fm/D_S}$ is employed to reduce the divergence between the feature representations of the prediction, $\hat{y}_i^{u/w}$ and the ground truth, $y_i^{l/w}$, encouraging to enhance segmentation quality on unannotated data. The second FM loss $L_{fm/D_P}$ is dedicated to reducing the difference in the feature statistics of the segmentation results under various data augmentation, $\hat{y}_i^{u/w}$, $\hat{y}_i^{u/s1}$, and $\hat{y}_i^{u/s2}$. The two FM losses are defined as follows:

$$L_{fm/D_S}(\theta_G) = \frac{1}{M}\sum_{i=1}^{M}L_{MAE}(D_S^k(y_i^{l/w}\copyright x_i^{l/w}), D_S^k(\hat{y}_i^{u/w}\copyright x_i^{u/w})), \quad (4)$$

$$L_{fm/D_P}(\theta_G) = \frac{1}{M}\sum_{i=1}^{M}[L_{MAE}(D_P^k(\hat{y}_i^{u/w}\copyright x_i^{u/w}), D_P^k(\hat{y}_i^{u/s1}\copyright x_i^{u/s1})) \\ + L_{MAE}(D_P^k(\hat{y}_i^{u/w}\copyright x_i^{u/w}), D_P^k(\hat{y}_i^{u/s2}\copyright x_i^{u/s2}))]/2, \quad (5)$$

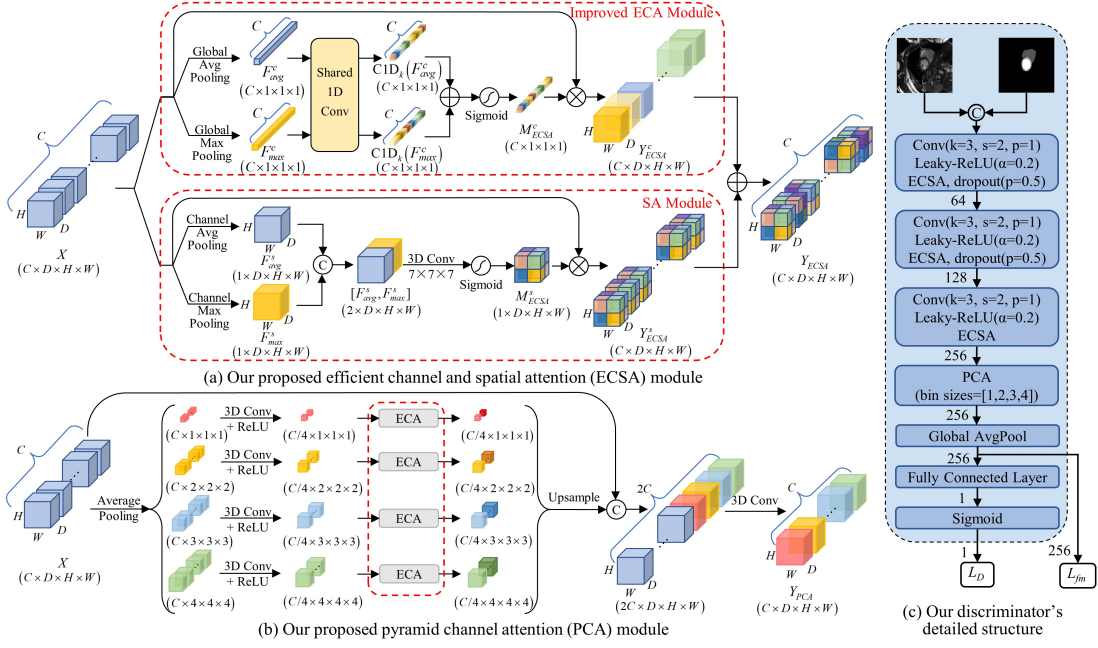

**Fig. 2.** Diagram of the 3D version of the proposed ECSA and PCA modules and our discriminator. (a) ECSA module. The ECSA module consists of our improved ECA module and SA module. (b) PCA module. The four improved ECA modules inside the red box are additions in our PCA module relative to the origin PPM module. (c) The discriminator network's detailed structure.

where $L_{MAE}$ is the Mean Absolute Error (MAE); $D_S^k$ and $D_P^k$ are intermediate representations of the discriminator networks $D_S$ and $D_P$ after the $k^{th}$ layer. Both discriminators take the concatenation of segmentation masks and their corresponding images as input. Note that the FM losses are only employed for the optimization of the segmentation network, whereas $D_S$ and $D_P$ only serve as feature extractors and do not maximize the loss $L_{fm/D_S}$ and $L_{fm/D_P}$.

*3.2.2. Efficient Spatial and Channel Attention*

The original squeeze-and-excitation (SE) block [30] was presented to incorporate a channel attention mechanism into the network. In the original SE block, spatial squeeze is achieved through a global average pooling operation, and channel excitation is achieved with a gating mechanism. However, the direct association between the channel and its weight is destroyed by the bottleneck [40]. Therefore, following the efficient channel attention (ECA) module proposed in [40], we avoid dimensionality reduction by conducting a 1D convolution for the channel excitation. Furthermore, since max pooling can provide another crucial clue about object features [41], we also gather spatial information by an additional max pooling operation. As illustrated in the upper part of Fig. 2 (a), for the input feature map $X \in \mathbb{R}^{C \times D \times H \times W}$, with $C$, $D$, $H$, and $W$ representing the channel, depth, height, and width dimensions, global spatial information is integrated into $F_{avg}^c, F_{max}^c \in \mathbb{R}^{C \times 1 \times 1 \times 1}$ through a global average pooling and a global max pooling. $F_{avg}^c$ and $F_{max}^c$ are input into a shared network, *i.e.*,

$$\begin{aligned} M_{ECSA}^c &= \sigma(\text{C1D}_k(AvgPool(X)) + \text{C1D}_k(MaxPool(X))) \\ &= \sigma(\text{C1D}_k(F_{avg}^c) + \text{C1D}_k(F_{max}^c)), \end{aligned} \quad (6)$$

where $\text{C1D}_k$ represents 1D convolution using a kernel size of $k$, which only involves $k$ parameters. This convolution operation is designed to capture interaction across channels, and the kernel size $k$ is the interaction coverage, which is adaptively determined based on channel dimension $C$:

$$k = \left| \frac{log_2(C)}{2} + \frac{1}{2} \right|_{odd}, \quad (7)$$

where $|t|_{odd}$ represents the closest odd number of $t$. The channel-wise refined output $Y_{ECSA}^c$ is computed by

$$Y_{ECSA}^c = M_{ECSA}^c(X) \otimes X, \quad (8)$$

where $\otimes$ represents element-wise multiplication, which first broadcasts the channel attention values along the spatial dimension.

Spatial attention (SA) is considered crucial for fine-grained image segmentation and complementary to channel attention. We leverage an SA module proposed in [41], shown in the lower section of Fig. 2 (a). We first employ max and average pooling across the channel dimension to highlight informative regions, generating two feature maps: $F_{max}^s, F_{avg}^s \in \mathbb{R}^{1 \times D \times H \times W}$. They are concatenated and input into a convolution layer, generating the spatial attention map $M_{ECSA}^s \in \mathbb{R}^{1 \times D \times H \times W}$, which is defined as follows:

$$\begin{aligned} M_{ECSA}^s(X) &= \sigma(f^{7 \times 7 \times 7}([AvgPool(X); MaxPool(X)])) \\ &= \sigma(f^{7 \times 7 \times 7}([F_{avg}^s; F_{max}^s])), \end{aligned} \quad (9)$$

where $f^{7 \times 7 \times 7}$ represents a 3D convolution using a filter size of $7 \times 7 \times 7$. The spatially refined output $Y_{ECSA}^s$ is computed by

$$Y_{ECSA}^s = M_{ECSA}^s(X) \otimes X, \quad (10)$$

where the spatial attention values are copied across the channel dimension while multiplying.

Lastly, we obtain the output $Y_{ECSA}$ of our ECSA module via element-wise addition of $Y_{ECSA}^c$ and $Y_{ECSA}^s$. Therefore, a location of the input feature map $X$ is assigned a high attention weight when both channel and spatial recalibration assign it a great significance. The ECSA modules are applied following each convolutional layer of the discriminator, promoting the network to learn feature maps of greater importance both channel-wise and spatially.

*3.2.3. Pyramid Channel Attention*

The deep neural network's receptive field size offers an estimate of the utilization of contextual information [42]. To expand the discriminator network's receptive field size, we employ the pyramid pooling module (PPM) proposed in [42] following the downsampling stage to establish a global contextual prior. Our PPM consists of four levels with bin sizes of $1 \times 1 \times 1$, $2 \times 2 \times 2$, $3 \times 3 \times 3$, and $4 \times 4 \times 4$. Following each pyramid level, a 3D convolution layer is employed to reduce the channel dimension to $1/4$ of the initial one. In order to emphasize significant channels and ignore less significant ones, we apply the improved ECA module proposed in Section 3.2.2 following each convolution layer. The reason why we do not utilize spatial attention is that, after average pooling, the feature map's depth, height, and width are very small. If we emphasize certain regions while ignoring others, it may result in the loss of multiple crucial features and undermine the network training. Finally, we upsample these feature maps and concatenate them with the original feature map, and employ a 3D convolution operation to decrease the channel dimension. The 3D version of our pyramid channel attention (PCA) module is shown in Fig. 2 (b). Our discriminator network's detailed structure is depicted in Fig. 2 (c), using *k*, *s*, and *p* to respectively represent convolutional kernel size, stride, and padding.

*3.3. Adaptive Self-Training*

In order to achieve entropy minimization [43], high-quality predicted results of unlabeled data are selected and transformed into hard pseudo-labels for supervised self-training [10]. In the FixMatch framework [10], the prediction result is used to generate a pseudo-label only when its largest class probability exceeds a fixed threshold. However, relying solely on the maximum predicted value often encounters limitations in accurately evaluating the prediction quality. Therefore, as depicted on the right side of Fig. 1, with the assistance of our adversarial learning, we use the output score of the discriminator $D_S$ as a confidence metric for the prediction's quality. Here, 0 should be allocated to the prediction results and 1 to the ground truth. High-quality predicted results selected are directly utilized for self-training with the self-training loss $L_{st/high}$ defined as:

$$L_{st/high}(\theta_G) = \begin{cases} [L_{CE}(\hat{y}_i^{u/s1}, \tilde{y}_i^{u/w}) + L_{Dice}(\hat{y}_i^{u/s1}, \tilde{y}_i^{u/w}) + \\ L_{CE}(\hat{y}_i^{u/s2}, \tilde{y}_i^{u/w}) + L_{Dice}(\hat{y}_i^{u/s2}, \tilde{y}_i^{u/w})]/4, & if\ D_S(\hat{y}_i^{u/w} \copyright x_i^{u/w}) > \gamma, \\ 0, & otherwise, \end{cases} \quad (11)$$

where $\gamma$ is the confidence threshold, determining the level of certainty required for a predicted result to be employed in self-training; $\tilde{y}_i^{u/w}$ is the pseudo-label produced from the predicted result $\hat{y}_i^{u/w}$ of weakly-augmented unlabeled data.

CutMix is an effective data augmentation strategy [44], which is often utilized as a strong perturbation within the weak-to-strong consistency regularization framework [45], [46]. Nevertheless, achieving a high semi-supervised performance is challenging since the pseudo-labels used to supervise the CutMixed image are low-quality [6]. To address this challenge, for predictions with confidence below the threshold $\gamma$, we employ bidirectional CutMix with labeled data to utilize them, rather than discarding them directly. Following the BCP [6], we first construct a zero-centered mask $M \in \{0,1\}^{H \times W \times D}$, where 1 and 0 respectively indicate whether the voxel belongs to the background or the foreground. The zero-value central region's size is $\beta H \times \beta W \times \beta D$, where $\beta$ is in the range of $(0,1)$. The bidirectional CutMix is defined as follows:

$$x_i^{m/s1} = \begin{cases} x_i^{l/w} \otimes M + x_i^{u/s1} \otimes (\mathbf{1} - M), & if\ \delta = 1, \\ x_i^{u/s1} \otimes M + x_i^{l/w} \otimes (\mathbf{1} - M), & if\ \delta = 0, \end{cases} \quad (12)$$

$$x_i^{m/s2} = \begin{cases} x_i^{l/w} \otimes M + x_i^{u/s2} \otimes (\mathbf{1} - M), & if\ \delta = 1, \\ x_i^{u/s2} \otimes M + x_i^{l/w} \otimes (\mathbf{1} - M), & if\ \delta = 0, \end{cases} \quad (13)$$

where $\mathbf{1} \in \{1\}^{H \times W \times D}$ and $\delta$ is a random variable with a value of 0 or 1. $x_i^{m/s1}$ and $x_i^{m/s2}$ are input into the segmentation network $G$, producing outputs $\hat{y}_i^{m/s1}$ and $\hat{y}_i^{m/s2}$.

In order to train the $G$, supervisory signals are also produced using bidirectional CutMix operations. For prediction of $x_i^{u/w}$ with confidence below the threshold $\gamma$, the initial pseudo-label $\tilde{y}_i^{u/w}$ is generated by taking the *argmax* operation on $\hat{y}_i^{u/w}$ and selecting the largest connected component, which will effectively eliminate outlier voxels. Then, the supervisory signal is obtained by bidirectional

CutMix between $\tilde{y}_i^{u/w}$ and ground truth $y_i^{l/w}$:

$$y_i^m = \begin{cases} y_i^{l/w} \otimes M + \tilde{y}_i^{u/w} \otimes (\mathbf{1}-M), & if\ \delta = 1, \\ \tilde{y}_i^{u/w} \otimes M + y_i^{l/w} \otimes (\mathbf{1}-M), & if\ \delta = 0, \end{cases} \quad (14)$$

$y_i^m$ will be used to supervise the $\hat{y}_i^{m/s1}$ and $\hat{y}_i^{m/s2}$, with the self-training loss $L_{st/low}$ expressed as follows:

$$L_{st/low}(\theta_G) = \begin{cases} [L_{CE}(\hat{y}_i^{m/s1}, y_i^m) + L_{Dice}(\hat{y}_i^{m/s1}, y_i^m) + \\ L_{CE}(\hat{y}_i^{m/s2}, y_i^m) + L_{Dice}(\hat{y}_i^{m/s2}, y_i^m)]/4, & if\ D_S(\hat{y}_i^{u/w} © x_i^{u/w}) \leq \gamma, \\ 0, & otherwise. \end{cases} \quad (15)$$

Mixed with the ground truth, the low-confidence pseudo-label's quality will be enhanced and then applied to self-training. Our adaptive self-training (AST) strategy can largely enhance the diversity of samples utilized for self-training while ensuring the pseudo-label's quality.

In summary, the final loss function for the segmentation network $G$ is expressed as:

$$L_G(\theta_G) = L_S + \lambda_{fm}(L_{fm/D_S} + L_{fm/D_P}) + \lambda_{st}(L_{st/high} + L_{st/low}), \quad (16)$$

where $\lambda_{fm}$, $\lambda_{st} > 0$ are the corresponding weights.

Algorithm 1 summarizes the overall procedure of our proposed AstMatch.

## 4. Experiments

### 4.1. Datasets and Pre-Processing

We evaluate our AstMatch on three medical image datasets:

(1) ACDC dataset: 2017 Automated Cardiac Diagnosis Challenge (ACDC) [15] dataset contains 100 cine magnetic resonance (MR) sequences. There are 3 classes to segment in every image, i.e., right ventricle (RV), myocardium (Myo), and left ventricle (LV). This dataset is employed to evaluate our method's performance in multi-class segmentation and 2D medical image segmentation tasks. Consistent with SS-Net [47] and BCP [6], we choose 2D U-Net [1] as the segmentation network and allocate 70, 10, and 20 scans for training, validation, and testing. The input patch size used for training is $256 \times 256$. We utilize 5% (3 scans) and 10% (7 scans) as annotated data, and the other data serve as unannotated data.

(2) LA dataset: Atrial Segmentation Challenge [16] dataset consists of 100 3D gadolinium-enhanced MR images with a resolution of $0.625 \times 0.625 \times 0.625$ mm³. Following the setting used

**Algorithm 1** The Training Process of AstMatch

**Training input**: labeled data $D_L = \{(x_i^l, y_i^l)\}_{i=1}^{N}$, unlabeled data $D_U = \{x_i^u\}_{i=N+1}^{N+M}$.

**Training output**: segmentation model $G$ parameterized by $\theta_G$; two independent discriminator networks $D_S$ and $D_P$ parameterized by $\theta_{D_S}$ and $\theta_{D_P}$, respectively.

1: Random initialize $\theta_G$, $\theta_{D_S}$, and $\theta_{D_P}$.
2: **for** i = 1: maximum iterations **do**
3:     //train the segmentation model $G$
4:     Apply weak augmentation to labeled data and unlabeled data → $(x_i^{l/w}, y_i^{l/w})$, $x_i^{u/w}$.
5:     Perform additional dual-stream strong augmentation on $x_i^{u/w}$ → $x_i^{u/s1}$, $x_i^{u/s2}$.
6:     Obtain the $G$ output: $\hat{y}_i^{l/w}$, $\hat{y}_i^{u/w}$, $\hat{y}_i^{u/s1}$, and $\hat{y}_i^{u/s2}$.
7:     Calculate $L_S$ (Equation. 1) on the labeled data for $G$ → $L_S$.
8:     Obtain the $D_S$ and $D_P$ outputs and calculate $L_{fm}$ (Equation. 4, 5) → $L_{fm/D_S}$, $L_{fm/D_P}$.
9:     **if** $D_S(\hat{y}_i^{u/w} \copyright x_i^{u/w}) > \gamma$
10:         Obtain the pseudo-label $\tilde{y}_i^{u/w}$ and calculate $L_{st/high}$ (Equation. 11) for $G$ → $L_{st/high}$.
11:     **else**
12:         Mix $x_i^{u/s1}$ and $x_i^{u/s2}$ with $x_i^{l/w}$ (Equation. 12, 13), and mix $\tilde{y}_i^{u/w}$ with $y_i^{l/w}$ (Equation. 14) → $x_i^{mixed\_s1}$, $x_i^{mixed\_s2}$, $y_i^{mixed}$.
13:         Input $x_i^{mixed\_s1}$ and $x_i^{mixed\_s2}$ into $G$ and obtain the output: $\hat{y}_i^{mixed\_s1}$, $\hat{y}_i^{mixed\_s2}$
14:         Calculate $L_{st/low}$ (Equation. 15) for $G$ → $L_{st/low}$.
15:     **end if**
16:     Calculate the total loss of $G$ (Equation. 16) → $L_G$.
17:     Update $\theta_G$ by optimizing $L_G$.
18:     //train the discriminator networks $D_S$ and $D_P$
19:     Obtain the $G$ output: $\hat{y}_i^{l/w}$, $\hat{y}_i^{u/w}$, $\hat{y}_i^{u/s1}$, and $\hat{y}_i^{u/s2}$.
20:     Obtain the $D_S$ and $D_P$ outputs and calculate $L_{D_S}$ (Equation. 2) and $L_{D_P}$ (Equation. 3) for $D_S$ and $D_P$ → $L_{D_S}$, $L_{D_P}$.
21:     Update $\theta_{D_S}$ by optimizing $L_{D_S}$. Update $\theta_{D_P}$ by optimizing $L_{D_P}$.

in BCP [6] and ASE-Net [7], the dataset is allocated with 80 scans for training and 20 for validation. The segmentation network is configured as 3D V-Net [2] and takes randomly cropped $112 \times 112 \times 80$ patches as input. We adopt labeled ratios of 5% (4 scans) and 10% (8 scans).

(3) Pancreas-NIH dataset: Pancreas-NIH [17] dataset is the most authoritative pancreas segmentation dataset, which contains 82 contrast-enhanced abdominal computed tomography (CT) volumes manually delineated. Every scan has a constant axial view resolution of $512 \times 512$ and the long axis length ranges from 181 to 466. We follow the data split in CoraNet [5] and BCP [6] and employ 3D V-Net as the segmentation network. We randomly crop $96 \times 96 \times 96$ patches as input during training and employ labeled ratios of 10% (6 scans) and 20% (12 scans).

*4.2. Implementation Details and Evaluation Metrics*

We conduct all experiments on an Nvidia GeForce RTX 4090 GPU. We choose the SGD optimizer to optimize the segmentation model, with the base learning rate, momentum, and weight decay set to 0.05, 0.9, and 5e-4. The discriminator networks are trained using the Adam optimizer, with the initial learning rate and betas set to 1e-4 and (0.9, 0.99). Similar to [48], the poly-learning policy is employed for all our networks, where the initial learning rate is scaled by a coefficient of $((1-\frac{iter}{max\_iter})^{0.9})$ in every iteration. The loss weights $\lambda_{fm}=0.1$, $\lambda_{st}=1.0$, and confidence threshold $\gamma=0.6$ are chosen as the default values. For the ACDC dataset, we configure the training iterations as 30k, batch size as 24, and $\beta^2$ as 0.3. For the LA and Pancreas-NIH datasets, we configure the training iterations as 15k, batch size as 8, and $\beta^3$ as 0.3. Equal amounts of annotated and unannotated data are included in each batch.

We use Dice similarity coefficient (DSC), Jaccard (JA), 95% Hausdroff distance (95HD), and average surface distance (ASD) to evaluate segmentation performance. DSC and JA primarily measure the degree of overlap between two regions, ranging from 0 to 1, which are more sensitive to the internal filling of segmentation. 95HD and ASD are surface-based metrics that are more sensitive to the boundary of segmentation result. Superior segmentation results are reflected in higher values of DSC and JA and lower values of 95HD and ASD.

*4.3. Comparative Experiments Results*

*4.3.1. Comparative Experiments on the ACDC Dataset*

We compare the proposed AstMatch with the supervised U-Net and six recent semi-supervised methods UA-MT [3], SASSNet [4], DTC [49], MC-Net+ [50], SS-Net [47], and BCP [6] on the ACDC dataset. For UA-MT [3], SASSNet [4], and DTC [49], we reproduce these methods specifically for 2D medical image segmentation. As for MC-Net+ [50], SS-Net [47], and BCP [6], since all the approaches adopt the identical configuration, we offer the optimal performance as reported in their original literature. Table 1 presents the quantitative results of various approaches on the ACDC test set with 5% and 10% labeled ratios. Compared to the fully supervised approach, all SSL approaches offer superior performance. The proposed AstMatch largely surpasses all SOTA methods. Utilizing the same 5% labeled

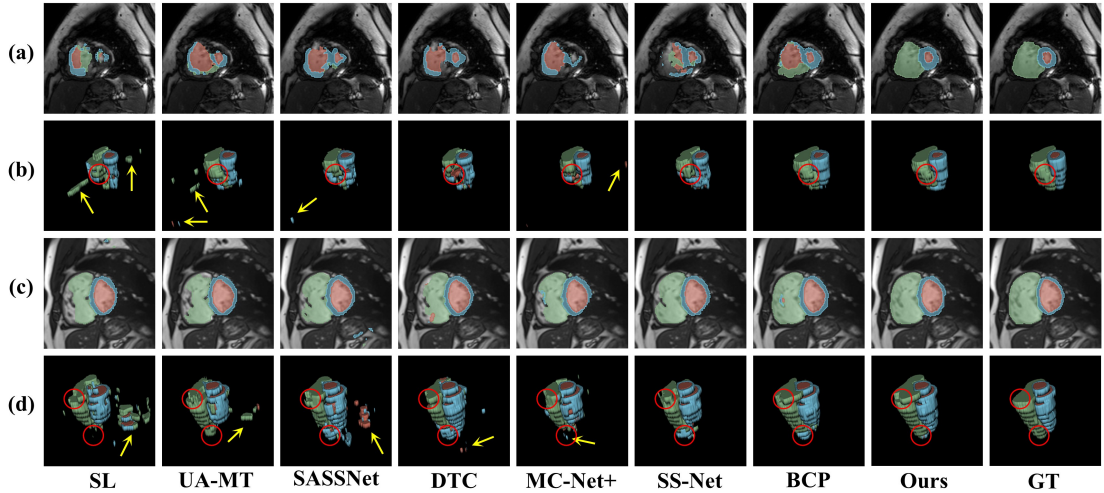

**Fig. 3.** Visualized results of various approaches on the ACDC test dataset trained with 5% labeled ratio. The rows (a) and (c) display prediction results for two specific slices, while the rows (b) and (d) present 3D visualizations for two cases.

**Table 1**

Comparisons with other approaches on the ACDC dataset with 5% and 10% labeled data. The optimal results are in bold, while the second-best results are underlined.

| Method | Scans used | Metrics | | | |
| --- | --- | --- | --- | --- | --- |
| | Labeled/Unlabeled | DSC(%)↑ | JA(%)↑ | 95HD(mm)↓ | ASD(mm)↓ |
| U-Net [1] | 3/0 | 52.33 | 42.03 | 26.99 | 11.72 |
| U-Net [1] | 7/0 | 80.75 | 69.72 | 9.68 | 2.56 |
| U-Net [1] | 70/0 | 91.66 | 84.94 | 4.33 | 0.90 |
| UA-MT [3] | 3/67 | 56.68 | 45.60 | 29.29 | 12.27 |
| SASSNet [4] | 3/67 | 64.29 | 52.21 | 22.42 | 7.90 |
| DTC [49] | 3/67 | 62.85 | 51.80 | 10.97 | 3.15 |
| MC-Net+ [50] | 3/67 | 64.50 | 53.92 | 11.21 | 3.82 |
| SS-Net [47] | 3/67 | 65.82 | 55.38 | 6.67 | 2.28 |
| BCP [6] | 3/67 | <u>87.59</u> | <u>78.67</u> | <u>1.90</u> | <u>0.67</u> |
| Ours | 3/67 | **89.12** | **81.00** | **1.68** | **0.52** |
| UA-MT [3] | 7/63 | 82.33 | 71.66 | 9.79 | 2.99 |
| SASSNet [4] | 7/63 | 82.35 | 71.33 | 9.28 | 2.54 |
| DTC [49] | 7/63 | 86.09 | 76.50 | 12.13 | 3.60 |
| MC-Net+ [50] | 7/63 | 87.10 | 78.06 | 6.68 | 2.00 |
| SS-Net [47] | 7/63 | 86.78 | 77.66 | 6.07 | 1.40 |
| BCP [6] | 7/63 | <u>88.84</u> | <u>80.62</u> | <u>3.98</u> | <u>1.17</u> |
| Ours | 7/63 | **90.47** | **83.15** | **1.62** | **0.49** |

data, AstMatch exhibits improvements compared to the second-ranked SSL method across all metrics (1.53% for DSC, 2.33% for JA, 0.22mm for 95HD, 0.15mm for ASD). Moreover, under 10% labeled data condition, our method obtains the highest DSC of 90.47% and JA of 83.15%. Compared with the U-Net trained with all 70 labeled data, our method has comparable DSC and JA and even lower 95HD

**Table 2**

Comparisons with other approaches on the LA dataset with 5% and 10% labeled data.

| Method | Scans used Labeled/Unlabeled | DSC(%)↑ | JA(%)↑ | 95HD(mm)↓ | ASD(mm)↓ |
|---|---|---|---|---|---|
| V-Net [2] | 4/0 | 39.27 | 31.12 | 49.36 | 9.78 |
| V-Net [2] | 8/0 | 80.64 | 69.95 | 15.40 | 2.67 |
| V-Net [2] | 80/0 | 91.63 | 84.60 | 5.61 | 1.51 |
| UA-MT [3] | 4/76 | 83.40 | 72.09 | 16.23 | 4.54 |
| SASSNet [4] | 4/76 | 83.82 | 72.74 | 15.44 | 4.53 |
| DTC [49] | 4/76 | 82.78 | 71.39 | 16.15 | 4.99 |
| MC-Net+ [50] | 4/76 | 84.50 | 73.65 | 13.86 | 3.13 |
| SS-Net [47] | 4/76 | 86.33 | 76.15 | 9.97 | 2.31 |
| BCP [6] | 4/76 | <u>88.02</u> | <u>78.72</u> | <u>7.90</u> | <u>2.15</u> |
| Ours | 4/76 | **90.30** | **82.38** | **5.66** | **1.76** |
| UA-MT [3] | 8/72 | 86.75 | 76.94 | 9.69 | 2.58 |
| SASSNet [4] | 8/72 | 87.32 | 77.72 | 9.62 | 2.55 |
| DTC [49] | 8/72 | 87.43 | 78.06 | 8.38 | 2.40 |
| MC-Net+ [50] | 8/72 | 88.90 | 80.17 | 8.16 | 1.82 |
| SS-Net [47] | 8/72 | 88.55 | 79.62 | 7.49 | 1.90 |
| BCP [6] | 8/72 | <u>89.62</u> | <u>81.31</u> | <u>6.81</u> | <u>1.76</u> |
| Ours | 8/72 | **91.09** | **83.70** | **6.02** | **1.49** |

and ASD with only 10% of the labeled data.

Fig. 3 illustrates some visual results on the ACDC test set trained with 5% labeled ratio. We do not perform any post-processing on these results to ensure fairness. It is observed that our AstMatch can obtain high-quality segmentation results with fewer isolated false predictions, as highlighted by the yellow arrows. One of the main reasons is that our AST strategy ensures the quality of pseudo-labels. Furthermore, our method achieves a more complete structure, highlighted by red circles. The key factor is that the discriminator networks with ECSA and PCA modules can feedback the segmentation quality on edges prone to being missed or misidentified. By optimizing the feature matching loss, our segmentation network is able to segment fine-grained details of the target organ.

*4.3.2. Comparative Experiments on the LA Dataset*

Table 2 presents the quantitative comparison results on the LA dataset with 5% and 10% labeled ratios. For a fair comparison, we utilize non-maximum suppression (NMS) as the post-process for all the approaches. For UA-MT [3], SASSNet [4], DTC [49], and MC-Net+ [50], since the original papers lacked experiments with a labeled ratio of 5%, we re-conducted experiments in the identical experimental setting. In summary, our AstMatch consistently outperforms other approaches across every metric under various labeled ratios. Our AstMatch method provides considerable improvements of more than 2.28% in DSC, 3.66% in JA, 2.24mm in 95HD, and 0.39mm in ASD over recent SSL approaches with 5% labeled ratio. It is also noticed that the AstMatch achieves 91.09% and 83.70% in DSC and JA with 10% labeled ratio, approaching the performance of V-Net trained with all 80 annotated data. As illustrated in Fig. 4, our AstMatch has fewer false negative predictions (row (b)) and false positive predictions (row (d)), highlighted by blue circles. Additionally, our predictions exhibit more a complete structure, attributed to the implementation of NMS to enhance the quality of pseudo-labels.

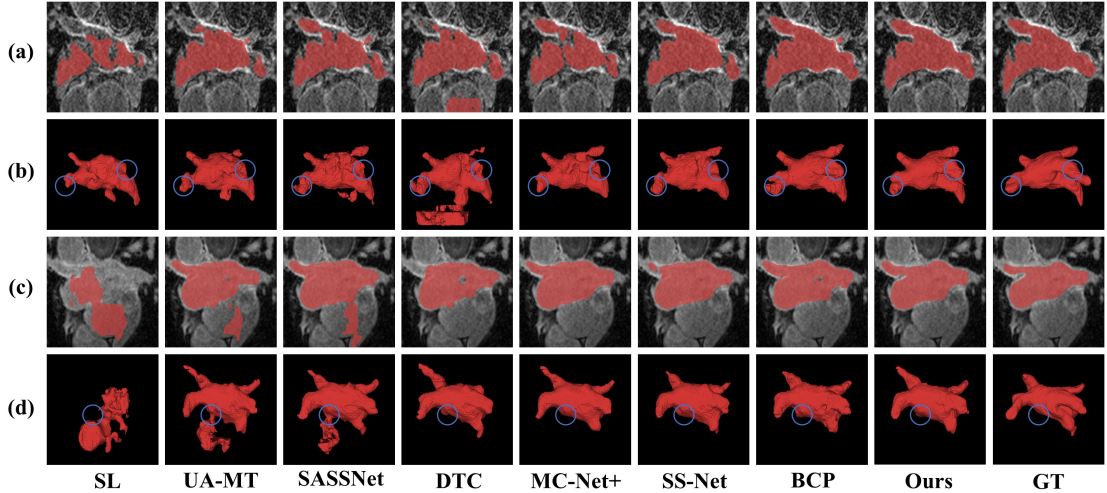

**Fig. 4.** Visualized results of various approaches on the LA validation dataset trained with 5% labeled ratio. The rows (a) and (c) display prediction results for two specific slices, while the rows (b) and (d) present 3D visualizations for two cases.

*4.3.3. Comparative Experiments on the Pancreas-NIH Dataset*

To further validate our proposed AstMatch, we performed comparative experiments on the Pancreas-NIH dataset. We conduct comparisons with 10% and 20% labeled ratios, and the results are shown in Table 3. We do not conduct any post-processing like NMS for fairness. Since we followed the data split in CoraNet [5], we reproduced the results for UA-MT [3], SASSNet [4], DTC [49], and MC-Net+ [50] in the identical experimental setting. For CoraNet [5] and BCP [6], we conducted experiments with 10% labeled ratio and reported their best performance with 20% labeled ratio presented in their original literature. Using 6 labeled data, our AstMatch maintains a considerable superiority over other approaches by more than 2.88% in DSC and 4.11% in JA. In the 20% setting, AstMatch achieves the highest 83.81% and 72.35% in DSC and JA, even exceeding the V-Net trained with all 62 labeled data

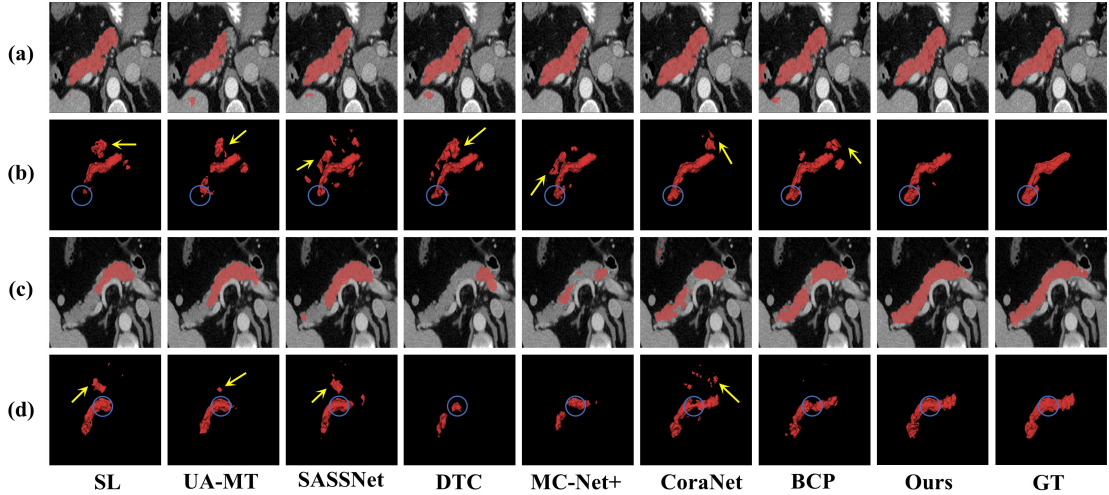

**Fig. 5.** Visualized results of various approaches on the Pancreas-NIH validation dataset trained with 10% labeled ratio. The rows (a) and (c) display predictions for two specific slices, while the rows (b) and (d) present 3D visualizations for two cases.

Table 3

Comparisons with other approaches on the Pancreas-NIH dataset with 10% and 20% labeled data.

| Method | Scans used | Metrics | | | |
|---|---|---|---|---|---|
| | Labeled/Unlabeled | DSC(%)↑ | JA(%)↑ | 95HD(mm)↓ | ASD(mm)↓ |
| V-Net [2] | 6/0 | 60.00 | 44.53 | 23.46 | 5.78 |
| V-Net [2] | 12/0 | 72.04 | 57.17 | 20.61 | 6.05 |
| V-Net [2] | 62/0 | 80.27 | 67.35 | 6.62 | 1.38 |
| UA-MT [3] | 6/56 | 64.33 | 48.95 | 21.46 | 6.72 |
| SASSNet [4] | | 64.69 | 49.48 | 23.78 | 8.16 |
| DTC [49] | | 61.60 | 46.24 | 23.40 | 6.39 |
| MC-Net+ [50] | | 66.17 | 50.64 | 26.18 | 7.87 |
| CoraNet [5] | | 68.13 | 53.20 | 16.92 | 5.55 |
| BCP [6] | | 80.87 | 68.16 | 8.26 | 2.63 |
| Ours | | **83.75** | **72.27** | **4.12** | **1.37** |
| UA-MT [3] | 12/50 | 73.06 | 58.80 | 17.62 | 5.12 |
| SASSNet [4] | | 74.70 | 61.85 | 12.30 | 3.78 |
| DTC [49] | | 75.38 | 62.11 | 8.27 | 2.98 |
| MC-Net+ [50] | | 77.13 | 63.65 | 11.67 | 2.73 |
| CoraNet [5] | | 79.67 | 66.69 | 7.59 | **1.89** |
| BCP [6] | | 82.91 | 70.97 | **6.43** | 2.25 |
| Ours | | **83.81** | **72.35** | 6.64 | 2.18 |

by 3.54% in DSC and 5.00% in JA. We also provide some visualized results of our AstMatch and other SSL approaches in Fig. 5. It is observed that our AstMatch can segment regions that are easily missed (rows (b) and (d)), marked by blue circles. Additionally, despite not employing any post-processing, our AstMatch still achieves a relatively complete structure, since our AST strategy ensures the quality of pseudo-labels.

**Table 4**

Ablation studies on the ACDC dataset with 5% labeled data.

| Method | | | Metrics | | | |
|---|---|---|---|---|---|---|
| ACR | *st/high* | *st/low* | DSC(%)↑ | JA(%)↑ | 95HD(mm)↓ | ASD(mm)↓ |
| | | | 52.33 | 42.03 | 26.99 | 11.72 |
| √ | | | 73.64 | 63.49 | 11.09 | 2.93 |
| √ | √ | | 75.13 | 64.02 | 12.17 | 4.39 |
| √ | | √ | 88.45 | 80.06 | **1.54** | **0.51** |
| √ | √ | √ | **89.12** | **81.00** | 1.68 | 0.52 |

**Table 5**

Ablation studies on the LA dataset with 5% labeled data.

| Method | | | Metrics | | | |
|---|---|---|---|---|---|---|
| ACR | *st/high* | *st/low* | DSC(%)↑ | JA(%)↑ | 95HD(mm)↓ | ASD(mm)↓ |
| | | | 39.27 | 31.12 | 49.36 | 9.78 |
| √ | | | 79.62 | 66.83 | 18.02 | 5.66 |
| √ | √ | | 83.62 | 72.52 | 15.44 | 4.86 |
| √ | | √ | 89.36 | 80.85 | 6.98 | 2.00 |
| √ | √ | √ | **90.30** | **82.38** | **5.66** | **1.76** |

*4.4. Ablation Studies*

*4.4.1. Ablation Studies on the ACDC Dataset*

To illustrate the contribution of each component, we conducted ablation studies on the ACDC dataset with 5% labeled ratio. Initially, we train a 2D U-Net using our supervision loss $L_S$. Based on the composition of the segmentation network's loss function $L_G$, we gradually incorporate submodules into the baseline model, including adversarial consistency regularization (ACR), self-training with pseudo-labels of high quality *st/high*, and self-training with pseudo-labels of low quality *st/low*.

As illustrated in Table 4, in comparison to the baseline model, the addition of ACR achieves the increase of DSC by 21.31%, proving the advantages of adversarial learning in extracting latent information. Our ACR is composed of two discriminators $D_S$ and $D_P$, whose roles are complementary. $D_S$ is mainly used for the feature consistency between predicted results and ground truth and the selection of high-quality pseudo-labels, whereas $D_P$ is employed for predictive consistency under perturbations with varying intensities. When ACR is employed, our AST strategy can be implemented based on the output scores from $D_S$. The addition of *st/low* offers greater improvement compared to *st/high*, as it incorporates bidirectional CutMix, introducing more variation in boundary regions of the segmentation network's input.

**Table 6**

Ablation studies on the Pancreas-NIH dataset with 10% labeled data.

| Method | | | Metrics | | | |
|---|---|---|---|---|---|---|
| ACR | *st/high* | *st/low* | DSC(%)↑ | JA(%)↑ | 95HD(mm)↓ | ASD(mm)↓ |
| | | | 60.00 | 44.53 | 23.46 | 5.78 |
| √ | | | 62.65 | 47.73 | 25.23 | 8.97 |
| √ | √ | | 66.84 | 51.99 | 21.11 | 6.56 |
| √ | | √ | <u>82.41</u> | <u>70.35</u> | <u>4.75</u> | <u>1.58</u> |
| √ | √ | √ | **83.75** | **72.27** | **4.12** | **1.37** |

*4.4.2. Ablation Studies on the LA Dataset*

Table 5 shows the results of ablation experiments on the LA validation set with 5% labeled ratio. The supervised V-Net achieves a DSC of 39.27%, and the introduction of our proposed ACR gains a 40.35% improvement in DSC. Furthermore, based on V-Net with ACR, the addition of *st/low* achieves a 5.74% more improvement in DSC than the addition of *st/high*, indicating the effectiveness of bidirectional CutMix. The optimal performance is obtained when employing both *st/high* and *st/low*. The main reason is that we supervise the segmentation mask of a complete unlabeled image in *st/high*, whereas in *st/low*, we supervise the segmentation result of a CutMixed image. Therefore, the utilization of both *st/high* and *st/low* enhances the diversity of images involved in the self-training process.

*4.4.3. Ablation Studies on the Pancreas-NIH Dataset*

Table 6 shows the results of ablation studies on the Pancreas-NIH dataset with 10% labeled ratio, which indicates the contribution and complementarity of our components. Compared to the supervised V-Net, the addition of ACR obtains an improvement of 2.65% for DSC. Based on ACR, the implementation of *st/high* and *st/low* achieve the increase of DSC by 4.19% and 19.76%, respectively, indicating the contribution of our proposed AST. Our method achieves the highest DSC of 83.75% when using ACR, *st/high*, and *st/low*, which demonstrates the complementarity of the submodules.

*4.5. Additional experiments*

*4.5.1. More Analysis on the Effectiveness of DusPerb*

With the self-training loss $L_{st}$ and feature matching loss $L_{fm/D_P}$, we utilize the segmentation results of weakly-augmented unlabeled data to supervise the predictions of their strongly-augmented counterparts. As depicted in Fig. 6, in the dual-stream perturbation (DusPerb) framework [37], two

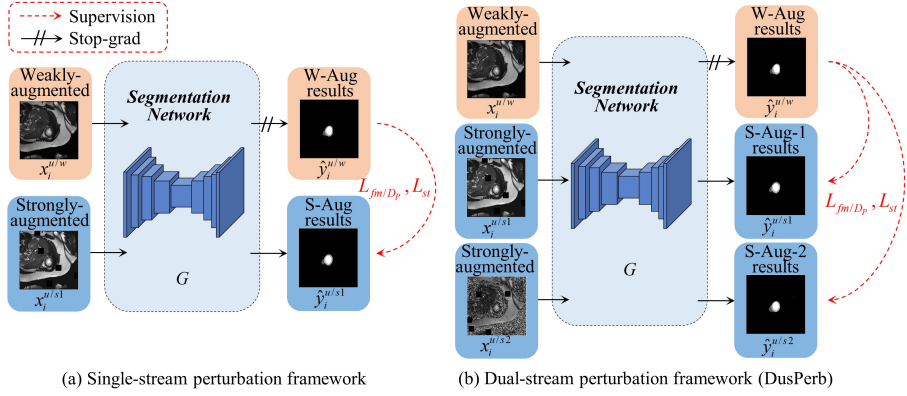

(a) Single-stream perturbation framework     (b) Dual-stream perturbation framework (DusPerb)

**Fig. 6.** (a) Single-stream perturbation framework. (b) Dual-stream perturbation framework (DusPerb). The red dashed curves represent supervision. The black lines with backslashes mean stop-gradient.

**Table 7**

Studies of DusPerb on the ACDC dataset with 5% labeled data.

| Method | Scans used | Metrics | | | |
|---|---|---|---|---|---|
| | Labeled/Unlabeled | DSC(%)↑ | JA(%)↑ | 95HD(mm)↓ | ASD(mm)↓ |
| Single-Stream | 3/67 | 88.58 | 80.26 | 1.98 | 0.64 |
| Dual-Stream | | **89.12** | **81.00** | **1.68** | **0.52** |
| Single-Stream | 7/63 | 90.19 | 82.62 | **1.55** | 0.55 |
| Dual-Stream | | **90.47** | **83.15** | 1.62 | **0.49** |

strongly-augmented versions are generated for each unlabeled image, whereas only one is generated in the single-stream perturbation framework. To further demonstrate the contribution of the DusPerb framework, we perform experiments on the ACDC dataset with 5% and 10% labeled ratios. As illustrated in Table 7, the DusPerb framework contributes to improvement in almost all the metrics compared to the single-stream framework, improving 0.54% for DSC and 0.74% for JA, and reducing 0.30mm for 95HD and 0.12mm for ASD under 5% labeled ratio. The key factor is that DusPerb [37] enhances the diversity of strongly-augmented unlabeled data with the randomness of RandAugment [38] and Cutout [39]. Moreover, regularizing both strong versions with an identical weak version could be considered as ensuring consistency between the both strong versions, which is extremely significant to model robustness [37].

*4.5.2. More Analysis on the Choice of GAN Loss*

In the adversarial training, we trained the segmentation network with the feature matching (FM) loss. To explain this setup, we evaluate the performance utilizing various GAN loss terms on the ACDC dataset with 5% labeled data. The standard GAN (SGAN) loss [22] in our experiments is defined as follows:

**Table 8**

Studies of different GAN loss terms of the ACDC dataset with 5% labeled data. SGAN: Standard GAN loss. FM: Feature matching loss.

| GAN Loss Terms | Metrics | | | |
|---|---|---|---|---|
| | DSC(%)↑ | JA(%)↑ | 95HD(mm)↓ | ASD(mm)↓ |
| SGAN only | 79.74 | 67.61 | 11.65 | 3.31 |
| FM only | **89.12** | **81.00** | **1.68** | **0.52** |
| SGAN + FM | 81.25 | 69.82 | 8.29 | 2.57 |

**Table 9**

Studies of different ST strategies on the ACDC dataset with 5% labeled data.

| ST Strategy | Metrics | | | |
|---|---|---|---|---|
| | DSC(%)↑ | JA(%)↑ | 95HD(mm)↓ | ASD(mm)↓ |
| w/o ST | 73.64 | 63.49 | 11.09 | 2.93 |
| ST in FixMatch | 85.30 | 75.36 | 7.55 | 2.32 |
| AST | **89.12** | **81.00** | **1.68** | **0.52** |

$$L_{SGAN/D_S}(\theta_G) = \frac{1}{M}\sum_{i=1}^{M} L_{BCE}(D_S(\hat{y}_i^{u/w} \copyright x_i^{u/w}), 1), \tag{17}$$

$$L_{SGAN/D_P}(\theta_G) = \frac{1}{M}\sum_{i=1}^{M}[L_{BCE}(D_P(\hat{y}_i^{u/s1} \copyright x_i^{u/s1}), 1) + L_{BCE}(D_P(\hat{y}_i^{u/s2} \copyright x_i^{u/s2}), 1)]/2, \tag{18}$$

$$L_{SGAN} = L_{SGAN/D_S} + L_{SGAN/D_P}, \tag{19}$$

where $L_{BCE}(\cdot)$ is binary-class cross-entropy loss. As illustrated in Table 8, the network trained with SGAN loss obtains the lowest 79.74% of DSC and 67.61% of JA. By combining the SGAN loss with the FM loss, the performance is improved, achieving improvements of 1.51% for DSC and 2.21% for JA. Furthermore, our AstMatch reaches 89.12% of Dice and 81.00% of JA when only using FM loss, indicating the beneficial influence of FM loss on the training stability of GAN and high-level consistency regularization.

*4.5.3. More Analysis on the Effectiveness of AST*

In order to illustrate the effectiveness of our AST, we compare the performance of our approach without self-training (ST), with ST in FixMatch [10], and with our AST on the ACDC dataset using 5% labeled data. As illustrated in Table 9, compared to the baseline without ST, our method with ST in FixMatch records improvements of 11.66% for DSC and 11.87% for JA, where the maximum class probability of each pixel in the predicted result determines whether to include it in the pseudo-label. The highest performance is achieved when equipping our method with AST. The main reason is that the confidence measure based on the discriminator scores can effectively evaluate the quality of a pre-diction. Furthermore, we apply bidirectional CutMix with labeled data to incorporate predictions with confidence below the threshold into the self-training strategy, enhancing sample diversity.

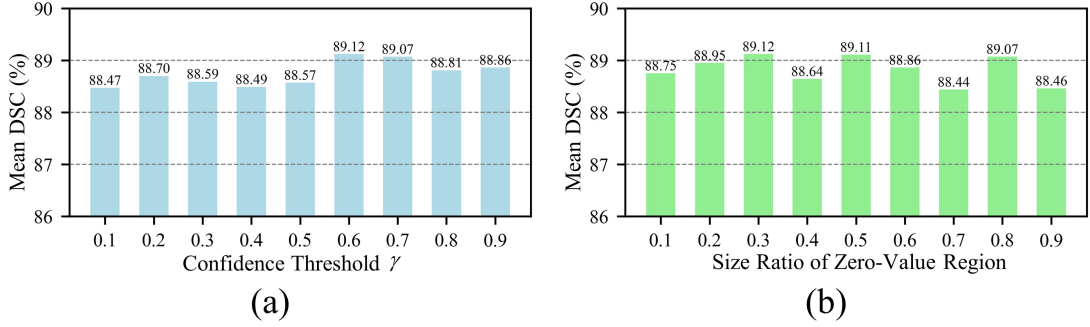

**Fig. 7.** DSC performances on the ACDC test dataset under various hyperparameter configurations with 5% labeled data. DSC performances with different (a) confidence thresholds $\gamma$, (b) size ratios $\beta^2$ of zero-value region in the CutMix mask.

*4.5.4. Impacts of Hyperparameters*

In Fig. 7, we provide the performance of various hyperparameter experiments associated with the AST on the ACDC dataset using 5% labeled data. We first offer the mean DSC under various confidence threshold configurations in Fig. 7 (a). Overall, regardless of the specific threshold values, our model consistently outperformed other methods (≤87.59% DSC). The DSC metrics remained consistently high across a broad range of confidence threshold values, indicating the robust effectiveness of our method. The crucial factor is that we incorporate predictions identified as both high-quality and low-quality in AST. When the number of predictions identified as high-quality increases, the quantity of predictions identified as low-quality decreases, and vice versa. Our AST strategy ensures the stability of the self-training process.

Moreover, we evaluate the size ratio $\beta^2$ of the zero-value region in the CutMix mask, which is employed for predictions with confidence below the threshold $\gamma$ in our AST strategy. As shown in Fig. 7 (b), the DSC performance appears to be consistent with changes in the size ratio, exhibiting a tiny difference (0.68% DSC) between the highest and the lowest DSC. If the size ratio $\beta^2$ is too small, the tiny CutMixed foreground has limited capacity to represent the semantic information of the target organ. Conversely, when $\beta^2$ is too large, the randomness of the foreground region is significantly reduced, subsequently decreasing the diversity of the boundary regions in the CutMixed image. Since

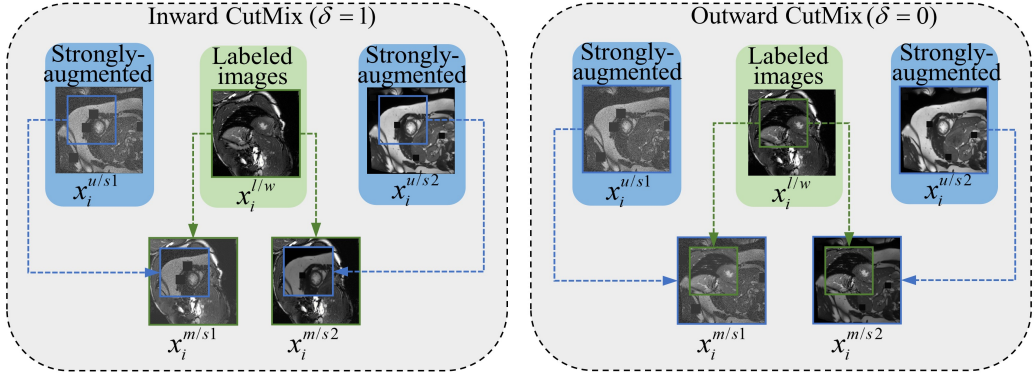

**Fig. 8.** Visualization of inward CutMix (left) and outward CutMix (right).

**Table 10**

Studies of different CutMix directions on the ACDC dataset with 5% labeled data.

| CutMix Direction | Metrics | | | |
|---|---|---|---|---|
| | DSC(%)↑ | JA(%)↑ | 95HD(mm)↓ | ASD(mm)↓ |
| In | 86.28 | 76.58 | 4.02 | 1.32 |
| Out | 85.79 | 75.94 | 5.98 | 1.59 |
| Ours | **89.12** | **81.00** | **1.68** | **0.52** |

we obtain the optimal performance when $\beta^2$ is set to 0.3, this value is employed as our final configuration. Note that when performing 3D medical image segmentation, the size ratio is set as $\beta^3 = 0.3$.

*4.5.5. More Analysis on the Choice of CutMix Direction*

As shown in Fig. 8, according to the variations in background and foreground selections, CutMix can be classified into inward CutMix and outward CutMix. To explore the impact of various CutMix directions in our AST, we performed three experiments on the ACDC dataset with 5% labeled data in Table 10. Inward CutMix and outward CutMix are denoted as "In" and "Out," referring to only using $x_i^{l/w} \otimes M + x_i^{u/s} \otimes (1-M)$ or $x_i^{u/s} \otimes M + x_i^{l/w} \otimes (1-M)$ for our AST. It is observed that both inward CutMix and outward CutMix exhibit inferior performances compared to the bidirectional CutMix, which treats labeled and unlabeled data uniformly, enabling greater semantic transfer between them.

*4.5.6. More Analysis on the Effectiveness of ECSA and PCA Modules*

To enhance the accuracy of the discriminator, we presented ECSA and PCA modules in Section 3.2.2 and incorporated them into the discriminator. To validate their contributions, we performed experiments on the LA dataset with 5% labeled data. As shown in Table 11, ECSA improves DSC by

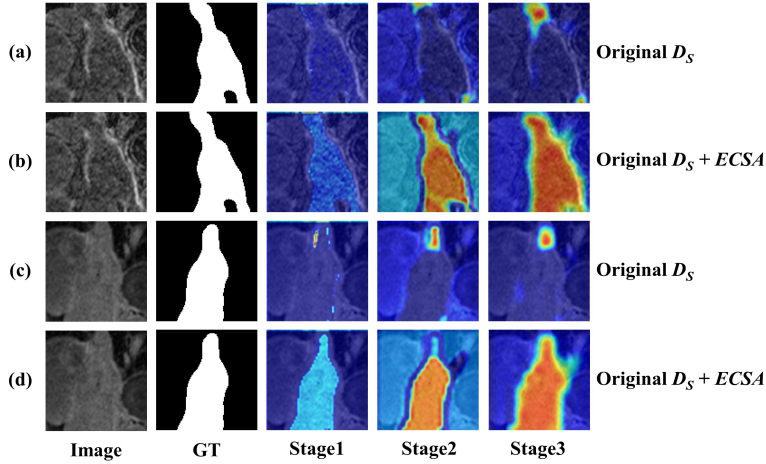

**Fig. 9.** Visualization of the class activation maps from different stages. Two instances are shown here. The rows (a) and (c) are class activation maps of the original discriminator without attention modules, while the rows (b) and (d) are class activation maps of the discriminator employing our ECSA modules. Stage1 to Stage3 represent maps from the first to the third convolutional layer, respectively.

**Table 11**

Studies of ECSA and PCA modules on the LA dataset with 5% labeled data.

| Method | | Metrics | | | |
|---|---|---|---|---|---|
| ECSA | PCA | DSC(%)↑ | JA(%)↑ | 95HD(mm)↓ | ASD(mm)↓ |
| | | 87.27 | 77.59 | 8.68 | 2.28 |
| √ | | 88.11 | 78.83 | 7.95 | 2.08 |
| | √ | 89.35 | 80.82 | 6.42 | 2.11 |
| √ | √ | **90.30** | **82.38** | **5.66** | **1.76** |

0.84% and JA by 1.24%, attributed to efficient channel and spatial recalibration. When the PCA module is incorporated into discriminators, DSC and JA improve by 2.08% and 3.23%, which demonstrates the effectiveness of introducing our improved ECA modules into the PPM. Our method achieves the highest DSC of 90.30% and JA of 82.38% when using both ECSA and PCA, indicating their complementary effects.

Additionally, the class activation maps from the segmentation discriminator $D_S$ with and without ECSA modules are illustrated in Fig. 9. We concatenated an image and its ground-truth segmentation mask into $D_S$. These maps are generated by the gradient-weighted class activation mapping (Grad-CAM) [51]. The rows (a) and (c) display class activation maps from the original discriminator without attention modules, and the rows (b) and (d) show class activation maps from the discriminator employing our ECSA modules. The discriminator has three convolutional layers, and we add an ECSA module after each convolutional layer. The red areas highlight the regions considered important for the

**Table 12**

Studies of different strong data augmentations on the Pancreas-NIH dataset with 10% labeled data.

| Strong Data Augmentations | | Metrics | | | |
|---|---|---|---|---|---|
| RandAugment | Cutout | DSC(%)↑ | JA(%)↑ | 95HD(mm)↓ | ASD(mm)↓ |
| √ | | <u>82.41</u> | <u>70.49</u> | <u>5.07</u> | <u>1.49</u> |
| | √ | 81.37 | 68.94 | 5.63 | 1.59 |
| √ | √ | **83.75** | **72.27** | **4.12** | **1.37** |

discriminator's output score. It can be observed that our proposed ECSA module enables the discriminator to focus more on the target organ region, effectively enhancing its ability to evaluate the quality of predicted segmentations.

*4.5.7. More Analysis on the Choice of Strong Data Augmentations*

Since the predicted results of weakly-augmented unlabeled data serve as supervisory signals for strongly-augmented versions, they must first undergo identical geometrical data augmentation. In order to produce appropriate prediction disagreement between them, we explore two additional strong data augmentations, *i.e.*, Cutout [39] and RandAugment [38] composed of intensity-based augmentations. In Table 12, we evaluate the effectiveness of each augmentation type and their combination on the Pancreas-NIH dataset with 5% labeled ratio. The results indicate that RandAugment is the most effective method for perturbing the unlabeled case. Furthermore, we observe that both RandAugment and Cutout are indispensable for achieving optimal performance; eliminating either one leads to inferior performance.

**5. Discussion and future work**

*5.1. Significance of the Proposed Algorithm*

According to the experimental results on three publicly available datasets, our AstMatch framework can be widely used for segmentation tasks involving various organs and imaging modalities. This can effectively reduce the workload of doctors performing manual delineation and enhance the efficiency and accuracy of medical image segmentation. Compared to fully supervised segmentation approaches that require a large quantity of annotated data for training, our proposed semi-supervised AstMatch framework achieves comparable segmentation accuracy with only a minimal amount of annotated training data. This significantly reduces the demand for large-scale annotated medical image datasets, which can greatly facilitate the application of automatic medical image segmentation systems in clini-cal practices.

*5.2. Limitations and Future Work*

In this paper, we propose ECSA and PCA modules to enhance the discriminator's ability to identify the quality of segmentation results. However, considering the distinctive shapes of specific organs, we can use more targeted approaches to improve the discriminator's ability. For instance, in the case of tubular organs like blood vessels, employing dynamic snake convolution [52] may further enhance discriminative performance. In the future, we plan to explore more targeted pseudo-label quality assessment methods and self-training strategies customized for the segmentation of different organs.

In addition, AstMatch does not incorporate extra modules or additional computational burden into the segmentation network, thereby enhancing its generalizability. Nevertheless, redesigning the segmentation network or incorporating advanced modules, such as transformer blocks [53], holds promise for more accurate segmentation, which is also our future research work.

## 6. Conclusion

In this paper, we proposed an adversarial self-training consistency framework named AstMatch for SSMIS. Firstly, the proposed ACR efficiently strengthens weak-to-strong consistency learning with two discriminators, which incorporate high-level consistency regularization. Then, two innovative attention modules ECSA and PCA are introduced to enhance the discriminator's ability to identify pseudo-labels with high quality. Furthermore, our proposed AST effectively ensures the quality of the pseudo-labels in self-training. Experiments on ACDC, LA, and Pancreas-NIH datasets indicate that our AstMatch outperforms cutting-edge SSL approaches and presents an effective solution for integrating the FixMatch framework into SSMIS. We expect our AstMatch can serve as a strong baseline in future SSMIS studies.

**CRediT authorship contribution statement**

**Guanghao Zhu:** Methodology; Software; Formal analysis; Investigation; Writing - original draft; Writing - review & editing; Visualization. **Jing Zhang:** Funding acquisition; Validation; Investigation; Writing - review & editing. **Juanxiu Liu:** Conceptualization; Validation; Writing - review & editing. **Xiaohui Du:** Investigation; Validation. **Ruqian Hao:** Validation; Visualization. **Yong Liu:** Supervision; Writing - review & editing. **Lin Liu:** Supervision; Resources; Writing - review & editing.

**Acknowledgement**

This work was funded in part by the Fundamental Research Funds for the Central Universities of China from the University of Electronic Science and Technology of China under Grant ZYGX2019J053 and Grant ZYGX2021YGCX020.